\documentclass[11pt]{article}
\usepackage{paspconf}
\usepackage{wrapfig}
\usepackage{graphicx}

\newcommand{\etal}{{\em et al\/}}
\newcommand{\kpc}{{\,\mathrm{kpc}}}
\newcommand{\kms}{{\,\mathrm{km/s}}}
\newcommand{\MHZ}{{\,\mathrm{MHz}}}
\newcommand{\MYM}{{\,\mu\mathrm{m}}}
\newcommand{\OmegaP}{{\Omega_{\mathrm{P}}}}

\newcommand{\xx}[1]{}

\begin{document}

\title{The Face-on View of the Milky Way: \\
       Gas Dynamics in the COBE NIR Bulge and Disk}
\author{Peter Englmaier}
\affil{University of Kentucky, Dep. of Physics \& Astronomy, USA,
ppe@pa.uky.edu}
\author{Ortwin Gerhard}
\affil{Astronomisches Institut der Universit\"at Basel, Switzerland,
gerhard@astro.unibas.ch}

\begin{abstract}
We report simulations of the gas flow in the gravitational potential
of the COBE NIR bulge and disk. These models 
lead to four--armed spiral structure between corotation of the bar and
the Sun, in agreement with the observed spiral arm tangents.
The 3-kpc-arm is identified with one of the arms emanating from the
ends of the bar.
\end{abstract}


\vspace*{-.5cm}
\section{Introduction}

The first face-on map of our Galaxy was constructed by Oort, Kerr \&
Westerhout (1958) from 21 cm observations, interpreting the observed
gas velocities in terms of circular motions in a thin disk. This map
revealed many arm--like features in the gas distribution. More recent
surveys of atomic hydrogen, molecular gas, HII regions, giant
molecular clouds (GMC), and other spiral arm tracers have helped to
constrain the location of Galactic spiral arms.  The majority of
observations appear to be consistent with a four armed spiral pattern
as suggested by Georgelin \& Georgelin (1976); see the review in
Vall\'ee (1995).

The gas in the inner Galaxy is not on circular orbits, however. This
is most evident from the so--called `forbidden' velocities which would
then imply gas in counterrotation. The most prominent example is the
3-kpc-arm, which appears in the $lv$-diagram at negative radial
velocities on both sides of the galactic center. In the direction of
the galactic center this arm is seen in absorption, which indicates
that it passes between the Sun and the Galactic center.

In recent years, near-IR observations with the COBE satellite have
motivated a series of new studies aiming to understand the dynamics of
the inner Galaxy. The NIR maps obtained with DIRBE
clearly show signatures of non-axisymmetric structure in the Galactic
bulge. Dwek \etal. (1995) used parametric bar models to determine the
basic properties of this triaxial bulge.

Full advantage of the observed asymmetries was taken by Binney,
Gerhard \& Spergel (1997; BGS), who applied a newly developed
Richardson--Lucy deprojection algorithm. By assuming that the bulge
has three mutually orthogonal planes of symmetry, they recovered
approximately the 3D distribution of NIR light in the Galactic center.
For their favored bar inclination of $\varphi=20\deg$, the bulge has
axis ratios 10:6:4 and semi--major axis $\sim 2\kpc$.  It is
surrounded by an elliptical disk that extends to $\sim 3.5\kpc$.
Outside the bar, the deprojected near IR luminosity distribution shows
a maximum $\sim 3\kpc$ down the minor axis.

Here we report gas dynamical simulations in this model and compare to
the observed spiral arm structure of the Galaxy. To follow the gas
flow we have used a two-dimensional (2D) `smoothed particle
hydrodynamics' (SPH) method, which has a large dynamical range in
resolution and can include self-gravity. We assume an isothermal
equation of state. See Englmaier \& Gerhard (1998; EG) for more
details.

\begin{table}
\small
\begin{tabular}{cccccp{7cm}}
\multicolumn{5}{c}{Spiral arm tangents in
longitude}&Measurement\\
\hline
29         & 50 & -50  & -32  &     &     HI, Weaver (1970)
Burton \& Shane (1970), Henderson (1977) \\
24, 30.5 & 49.5 & -50 & -30 &     & integrated $^{12}$CO, Cohen \etal. (1980), Grabelsky \etal. (1987) \\
25, 32 & 51  &     &     &     & $^{12}$CO clouds, Dame \etal. (1986) \\
25, 30      & 49  &     &     &     & warm CO clouds, Solomon \etal.
(1985)\\
24, 30      & 47 & -55 & -28 &     & HII-Regions (H109-$\alpha$), Lockman (1979), Downes \etal. (1980) \\
32         & 46 & -50 & -35 &     & $^{26}$Al, Chen \etal. (1996)\\
32      & 48  &-50,-58& -32 & -21 & Radio $408\,\MHZ$, Beuermann \etal.
(1985)\\
29         &    &     & -28 & -21 & $2.4\,\MYM$, Hayakawa \etal. (1981)
\\
26         &    & -47 & -31 & -20 & $60\,\MYM$, Bloemen \etal. (1990)
\\
\hline
30   & 49  & -51 & -31 & -21 & adopted mean\\
\hline
$\sim25$& 54& -44 & -33 & -20 & Model without halo\\
$\sim30$& 50& -46 & -33 & -20 & Model with halo $v_0=200\kms$ \\
\hline
\end{tabular}
\end{table}

\vspace*{-.5cm}
\section{Large scale morphology}

\begin{wrapfigure}{r}{8.0cm}
\includegraphics[trim=0 20 0 20,width=8.0cm]{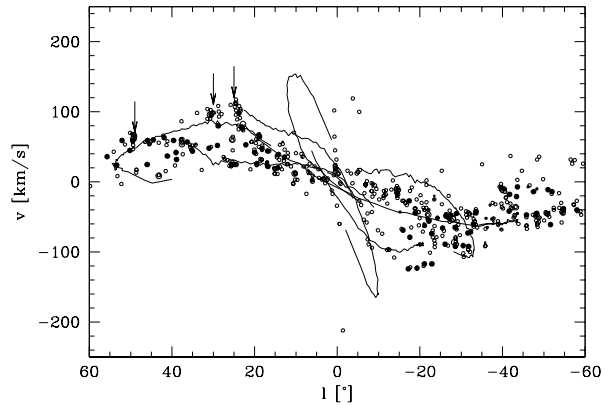}
\caption{HII regions ($\circ$) and GMC's ($\bullet$) compared to
model arms in the $lv$-diagram. See EG for references.}
\end{wrapfigure}
Here we describe the gas flow in the deprojected mass model of BGS for
$\varphi=20\deg$, and assume that the potential rotates with a
constant pattern speed $\OmegaP$, such that the corotation radius
falls between the molecular ring and the 3-kpc-arm at $3.4\kpc$. The
morphology of the gas flow (Fig.~2) is insensitive to the choice of
these parameters within reasonable bounds; see EG.

In these models a four armed spiral pattern forms, driven by the
rotating triaxial bulge and elliptical disk, and the strong mass
concentrations on the bar minor axis with negative quadrupole
moment. The latter are presumably signatures of the spiral arm heads
of the arms embedded in the molecular ring at $4\kpc$ radius. The
model arm tangents can be identified with the five observed arm
tangents (see Table).

The model arms also give a reasonable representation of the locations
of HII regions and GMC's in the $lv$-diagram (see Fig.~1). Finally,
the model gives a good approximation for the terminal velocity curve
(TVC), and hence the Galactic rotation curve, for $R \leq 5\kpc$
without including a dark halo component. The observed TVC is used to
fix the unknown velocity scale of the model, assuming a distance to
the galactic center $R_0=8\kpc$ and an LSR velocity $V_0=208\,\kms$.

\section{3-kpc arm and counter arm}
One spiral arm in the model qualitatively corresponds to the
3-kpc-arm.  The tangent of the model arm is at the correct longitude,
however, the non-circular motion at $l=0$ is somewhat smaller than for
the observed 3-kpc-arm. It has always been a puzzle why we do not see
a symmetric counter arm for the 3-kpc-arm at the far side of the
galaxy. The counter arm in our model runs almost parallel in the
$lv$-diagram to another arm which ends at about $30\deg$. In fact, the
observed arm there is known to split into two parts; see the two
concentrations of warm CO clouds at $25\deg$ and $30\deg$, indicating the
presence of two shocks (Solomon \etal. 1985; arrows in Fig.~1).

\section{Effect of spiral arm gravity}
Our model does not include a live stellar disk. Thus the gravitational
force of the stellar spiral arms is not included.  However, we can use
the gaseous arms as tracers to estimate the influence of stellar
spiral arms: We take a fraction of the mass from the stellar
background disk model and add it to the mass in gas particles. This
extra mass does not, however, enter the hydrodynamical equations. Then
we smooth the potential of the extra mass over
$\epsilon=1\kpc$, to mimic the fact that the stellar arms are much
broader than gaseous arms.

The resulting gas flow is similar to the previous one, in particular,
the spiral arm tangents are hardly changed.  The most significant
difference is that now the 3-kpc-arm displays just the right amount of
non-circular motion. In nature, gaseous spiral arms are driven by
stellar spiral arms. In our model, however, both are driven by the bar
and the non-axisymmetric features in the disk.

\section{Comparison to GMC's and HII regions}

\begin{figure}
\includegraphics[trim=0 10 0 20,width=8.7cm]{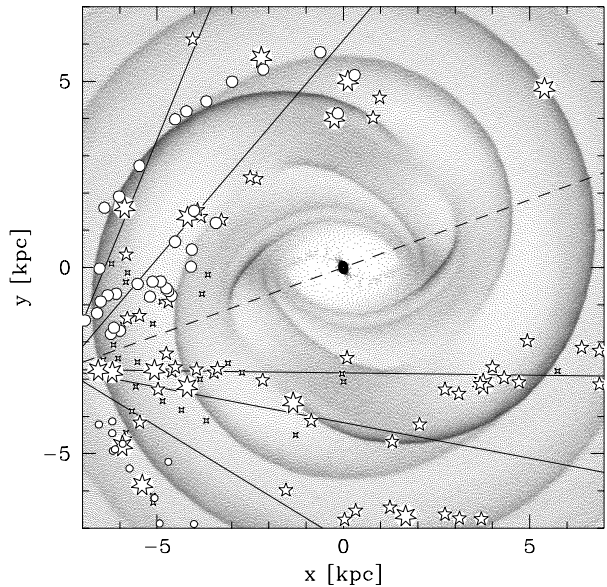}%
\vskip -5mm
\caption{Gas model face-on compared to HII ($\star$) and GMC ($\circ$)
and observed tangents (lines). The bar major axis is horizontal,
and the Sun is at $x=-7.5\kpc$, $y=-2.7\kpc$.
}
\end{figure}

The idea of a four armed spiral pattern goes back to Georgelin \&
Georgelin (1976), who used HII regions as tracers for spiral arms.  In
Fig.~2 we compare the face-on view of our model to the tracer
positions inferred from their and other more recent studies
including surveys of GMC's (see EG for references).  Note that this
plot does not take into account the non-circular motions of the
tracers. Correcting for this effect will focus the points towards the
spiral arms, due to velocity crowding; thus the correspondence will
improve.

\acknowledgments

This work was supported by the Swiss NSF grants 21-40'464.94 and
20-43'218.95, and the NASA grants WKU-522762-98-6 and NAG5-3841.

\end{document}